# Self-compression of ultrahigh-peak-power lasers

Renjing Chen[1, 2, 3], Wenhai Liang[1, 2], Yilin Xu[1, 2], Xiong Shen[3],

Peng Wang[3], Jun Liu[1, 2, 3]*, Ruxin Li[1, 2, 3]

[1]*State Key Laboratory of High Field Laser Physics and CAS Center for Excellence in Ultra-intense Laser Science, Shanghai Institute of Optics and Fine Mechanics, Chinese Academy of Sciences, Shanghai 201800, China*

[2]*University Center of Materials Science and Optoelectronics Engineering, University of Chinese Academy of Sciences, Beijing 100049, China*

[3]*Zhangjiang Laboratory, 100 Haike Road, Pudong, Shanghai 201210, China*

*Corresponding author: jliu@siom.ac.cn*

**Abstract:** Pulse self-compression is a simple and economical method for improving the peak power of ultra-intense laser pulses. By solving a modified nonlinear Schrodinger equation considering the fifth-order susceptibility, we found that self-compression appeared even in normally dispersive medium owing to the negative fifth-order susceptibility inducing a mass of negative frequency chirp. Furthermore, negatively pre-chirped pulses allow for self-compression at lower intensity, avoiding medium damage. We numerically analyze the optimal choice of pre-chirp, input intensity, and medium length. A proof-of-principle experiment successfully proves the above theoretical findings. It is expected that petawatt or even exawatt laser pulses with 25 fs/15 fs transform limited pulse duration can be self-compressed to about 9.9 fs/7.6 fs in normally dispersive medium, such as fused silica glass plate.

## 1. Introduction

Ultrashort pulse duration and ultrahigh peak-power are two key properties of femtosecond laser pulses. The feature of ultrashort pulse duration is highly desirable in a wide variety of research domains ranging from femtosecond chemistry, ultrafast spectroscopy to nonlinear optical microscopy, as a key requirement. The peculiarity of ultrahigh-peak-power, which is directly associated with the pulse duration, had been extensively employed in ultrahigh-intensity laser physics. For a given pulse energy, a shorter pulse duration corresponds to a higher peak power. Consequently, the compression of pulses is consistently crucial for femtosecond laser applications.

In the past several decades, pulse compression was mainly based on two conjoint processes: spectral broadening achieved through self-phase modulation (SPM) process in normally dispersive medium, followed by spectral dispersion compensating through introducing negative dispersion[1-11]. Various medium including bulk medium[1-3], fiber[4, 5], gas-filled hollow fiber[6, 7], multiple thin plates[8, 9] and multiple pass cell[10, 11] had been used to broaden the laser spectrum. Additionally, compensation for the induced positive dispersion during SPM processes[1-11] was achieved by prism pair or chirped mirrors. In the case of abnormally dispersive region, self-compression is an effective method which had been observed in many works[12-18], with their laser spectra usually in the near infrared or mid-infrared spectral range. The self-compression processes inherently compensate for the SPM induced positive dispersion automatically through the negative dispersion induced by material. Consequently the setup of self-compression is remarkably straightforward, as there is no additional dispersive compensation process.

Recently, to improve an ultrahigh-peak-power laser economically and simply, pulse compression methods had been extended to 100s terawatt(TW) or petawatt(PW) lasers experimentally[2, 19] and even exawatt(EW) laser in theory recently[20]. However, all the central wavelengths of PW laser pulses were less or around 1 um, which located at normally dispersive region. Therefore, chirped mirrors with several hundreds of millimeters utilized to compensate the induced dispersion[20] greatly limit the further compression of PW laser pulses, for their high cost or even unrealizability.

As an alternative approach, self-compression, launched a new direction in post-compression of ultrahigh-peak-power laser pulse because of advantage of not requiring chirped mirrors. Few previous experimental and theoretical works had been reported the possibility[21]. Self-compression has been successfully observed and demonstrated in gas based filamentation[22, 23], where the SPM induced dispersion can be compensated by negative nonlinear refractive index results from multi-photon ionization during the propagation. Besides gas based filamentation, a 75 fs negatively chirped pulse was self-compressed to 27 fs in normally dispersive solid medium, which was approximately half the duration of the 50 fs transform limited (TL) input duration[24]. However, the underlying mechanism responsible for the self-compression process in normally dispersive medium remains unclear.

In this manuscript, we demonstrate both numerically and experimentally that fifth-order nonlinear susceptibility and the negative pre-chirp can contribute to self-compression, and allows shorter pulse duration and higher peak power for ultra-intense laser installations. By solving a modified nonlinear Schrodinger equation, the results clearly demonstrate that the fifth-order nonlinear susceptibility plays a key role in achieving self-compression in normally dispersive medium. Moreover, appropriate negative pre-chirp can effectively reduce the peak input intensity required for self-compression, otherwise, it may damage the medium before the appearance of self-compression. As a result, the mechanism of self-compression in normally dispersive medium can be attributed to the combination of the fifth-order nonlinear susceptibility and the negative pre-chirp effectively balancing the induced positive dispersion. This theoretical conclusion has been experimentally validated through a proof-of-principle experiment conducted on fused silica glass plates. Together with asymmetrical four-grating system involving beam smoothing effect proposed in previous works[25, 26], the small-scale self-focusing (SSSF) effect can be well restrained or eliminated. Therefore, the self-compression technique can be implemented in future PW or EW ultrahigh-peak-power laser facilities without requiring any large chirp mirrors.

## 2. Numerical simulation

The inclusion of fifth-order nonlinear susceptibility $\chi^{(5)}$ in our propagation model accounts for the negative second-order nonlinear refractive index $n_4$ in glass plates[27] and other medium[28-31]. Then, previous nonlinear Schrodinger equation[32] can be modified as:

$$\frac{\partial E}{\partial z} = \frac{i}{2k_0} \nabla_\perp^2 E + \frac{i\beta_2}{2} \frac{\partial^2 E}{\partial T^2} + ik_0 \frac{\Delta n}{n_0} E - \frac{n_2}{c} \frac{\partial}{\partial T} |E|^2 E - \frac{n_4}{c} \frac{\partial}{\partial T} |E|^4 E - \frac{\alpha^{(K)}}{2} |E|^{2K-2} E \quad (1).$$

Please refer to Supplemental 1 for a detailed description of Equation (1). The input pulse with the pre-chirped parameter $C$ can be expressed as: $E = \sqrt{I_0} \exp(-r^2/r_0^2) \times \exp(-1 - iC/2)(t/T_0)^2$, where a negative value of $C$ indicates pulses with negative pre-chirp. The validity of our modified numerical simulation model is demonstrated through a proof-of-model simulation, which utilizes the same input optical parameters as the previous experiment[24]. Please refer to Supplemental 2 for the detail.

## 3. Results

### 3.1 Fifth-order susceptibility

Considering the negative fifth-order susceptibility, the induced nonlinear refractive index change can be expressed as $\Delta n = n_2 I + n_4 I^2 = (n_2 + n_4 I)I$ [31]. When $n_2 + n_4 I < 0$ or $I > |n_2/n_4|$, a negative rather than positive nonlinear refractive index is induced. For fused silica, the measured value of $n_2$ at 800 nm wavelength and room temperature is $3.20 \times 10^{-4}$ cm$^2$/TW[33], while for a wavelength of 438 nm, the measured value of $n_4$ is $-8.25 \times 10^{-4}$ cm$^4$/TW$^2$[27]. Based on the relationship among $n_4$, $n_2$ and wavelength[28], it can be calculated that at 800 nm in fused silica, $n_4$ is approximately $-4.4 \times 10^{-4}$ cm$^4$/TW$^2$. This implies that a negative nonlinear refractive index can be induced when the intensity exceeds 0.73 TW/cm$^2$. It should be noted that self-compression can only occur when the induced negative nonlinear refractive index is sufficiently large to broaden the spectrum and compensate for material-induced positive dispersion simultaneously.

The induced new frequency (INF) by SPM in the medium can be expressed as

$$\omega(r, z, T) = \omega_0 - \frac{\partial}{\partial T}\Delta\varphi(r, z, T) \quad (2).$$

Here, the phase change after propagating through a length of medium, $\Delta L$, can be represented as $\Delta\varphi(r, z, T) = k_0[n_2 I(r, z, T) + n_4 I^2(r, z, T)]\Delta L$. The temporal evolution of INF curves at different input intensities reveals the critical input intensity required for achieving self-compression. As an illustrative example, Fig. 1 presents the INF without and with the fifth-order susceptibility at different input intensities for a unit thickness of 1 um in fused silica glass plate and a 50 fs input pulse.

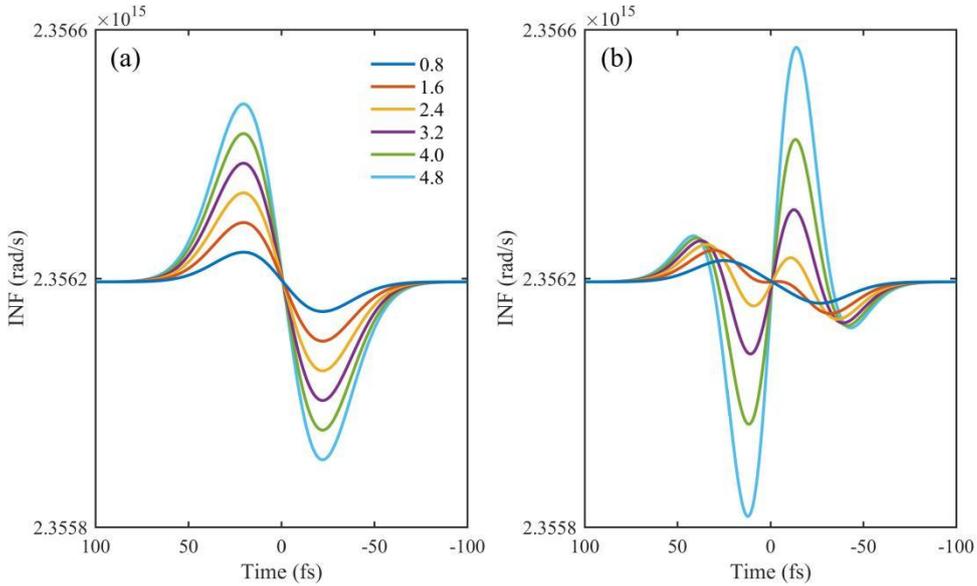

Fig. 1. The evolution of INF depending on time at different input intensities from 0.8 to 4.8 TW/cm$^2$ (a) without and (b) with the fifth-order susceptibility.

In the absence of the fifth-order susceptibility $\chi^{(5)}$, positive chirp is consistently induced, as depicted in Fig. 1(a). However, when taking into account $\chi^{(5)}$, the increase in negative frequency chirp surpasses that of induced positive frequency chirp with increasing input intensity, as illustrated in Fig.

1(b). This evolution of INF is in accordance well with the previous Z-scan research findings on high-order nonlinearity[31]. It should be noted that the multi-photon absorption effects have been considered in this work[34]. At an approximately 2.4 TW/cm$^2$ input intensity, the induced positive and negative frequency chirps are balanced, which means there is no spectral broadening but pulse broadening due to the material dispersion theoretically. As the input intensity keep increasing, the spectra will be broadened again by the total negative frequency chirp. However, due to relatively large positive material dispersion, the pulse will still be broadened at beginning. Only at a relatively high input intensity does a point arises where spectral broadening and induced negative frequency chirp counterbalance material dispersion, leading to self-compression finally. With further increases in input intensity, the self-compressed pulse become progressively shorter and shorter, before eventually undergoing splitting. Note that the medium damage threshold is not considered here.

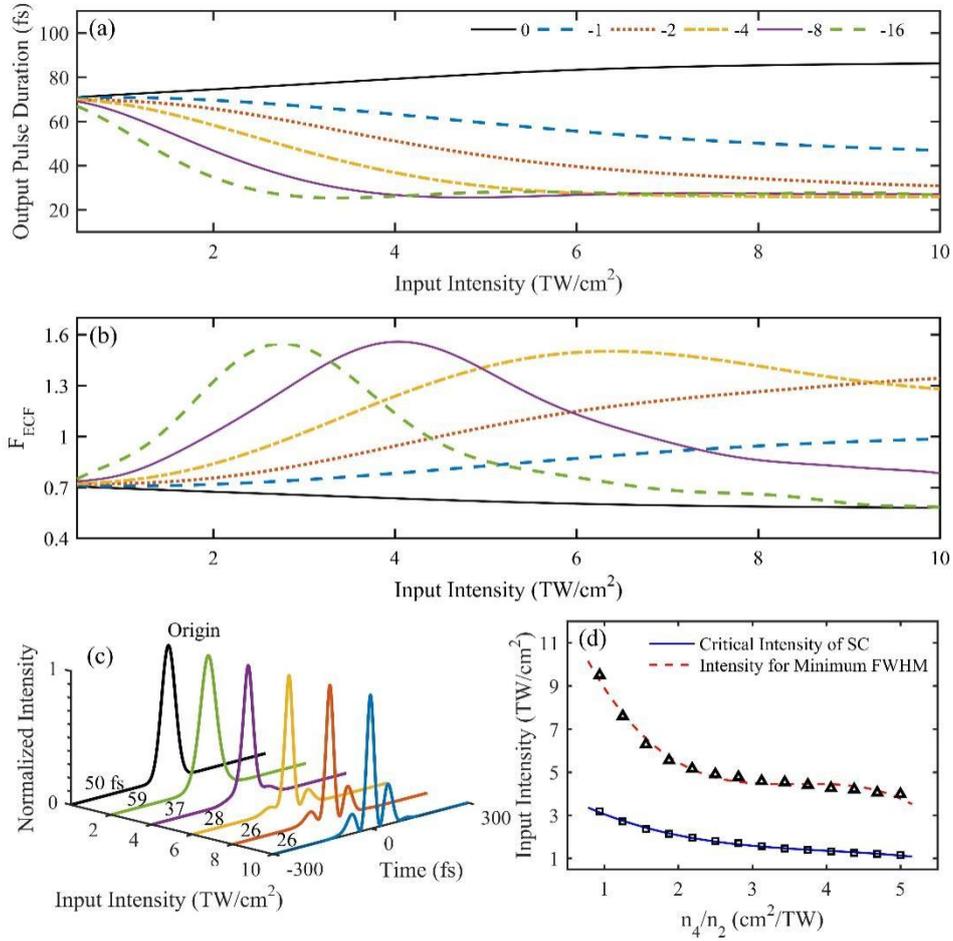

Fig. 2. (a) and (b) The output duration and $F_{ECF}$ curves changing with the input intensities. (c) The output normalized temporal profiles in linear coordinate with $n_4$=-4×10$^{-4}$ cm$^4$/TW$^2$. (d) The fitted curves of critical input intensities achieving self-compression or the minimum duration at different $n_4/n_2$.

To thoroughly investigate the contribution of $\chi^{(5)}$ to the self-compression, we conducted simulation using a 3-mm-thick fused silica plate and a 50 fs TL Gaussian pulse. The input intensities were varied from 0.5 to 10.0 TW/cm$^2$. When $\chi^{(5)}$ was not considered, the positive chirp induced by SPM could not be self-compensated by the positive material dispersion, resulting in continuous

4 / 19

broadening of the output pulse duration without any self-compression (as depicted by the black-solid line in Fig. 2(a)). By varying $n_4$ from -1×10$^{-4}$ to -16×10$^{-4}$ cm$^4$/TW$^2$, the output duration of the main pulse was compressed to nearly identical shortest duration but at different speeds, disregarding the damage of the medium. Although these self-compression processes seemed stable, satellite pulses emerged on both sides at higher input intensities. For $n_4$=-4×10$^{-4}$ cm$^4$/TW$^2$, as shown in Fig. 2(c), the output duration gradually decreased from 59 fs to 26 fs as the input intensities increasing from 2.0 to 8.0 TW/cm$^2$. Upon reaching an input intensity of 10 TW/cm$^2$, the pulse became noticeably split due to the high sidebands. To eliminate the impact of sidebands, a new parameter, namely pulse quality $Q = \frac{\int_{-a}^{a}|I_{main}|dT}{\int_{-\infty}^{+\infty}|I_{all}|dT}$, defined as the ratio of the central main pulse energy to the total pulse energy[35, 36], is proposed to precisely evaluate the output pulses quality, where the $I_{main}$ is the intensity of center main pulse without sidebands. Additionally, we introduce an effective compression factor $F_{ECF}$ to qualify the actual compression quantity. $F_{ECF}$ is calculated as $F_{ECF}$=CF×$Q_{out}$/$Q_{TL\_in}$, where CF represents the traditional compression factor defined as the ratio between input and output main pulse durations, while $Q_{out}$ and $Q_{TL\_in}$ denote the pulse qualities of the output pulse and the input TL pulse respectively. It should be noted that all simulated input pulses are ideal Gaussian pulses without any sidebands. In comparison to the output duration shown in Fig. 2(a), the effective compression factor $F_{ECF}$ gradually increased and reached nearly identical maximum values for different nonzero $n_4$ at different input intensity, as depicted in Fig. 2(b). Additionally, Fig. 2(b) demonstrates a relatively stable $F_{ECF}$ region at the top of each curve, where $F_{ECF}$ remains insensitive to fluctuation of input intensity. This indicates the possibility of achieving stable self-compression. Furthermore, Fig. 2(d) provides critical input intensities required for self-compression or obtaining the minimum output durations in a 3-mm-thick fused silica as reference points, revealing that higher absolute $n_4$ values correspond to lower and safer input intensities.

According to the aforementioned results, it can be inferred that $\chi^{(5)}$ or $n_4$ is an essential prerequisite for achieving self-compression in normally dispersive region. With an increasing absolute value of $n_4$ while keeping $n_2$ fixed, the attained shortest pulse duration remains nearly constant; however, self-compression appears at relatively lower input intensity levels. Therefore, $|n_4/n_2|$ serves as a crucial parameter for safely accomplishing self-compression in normally dispersive medium. It should be noted that employing a higher $|n_4/n_2|$ value and a lower input intensity facilitates easier and safer achievement of self-compression while considering material threshold damage.

However, self-compressions were not found in previous works on post pulse compression with glass plates. This is primarily due to the relatively high input intensity required to achieve self-compression using currently employed solid media. Previous experiments on post-compression utilizing SPM and chirp mirror compensating typically operated at input intensities lower than 3.0 TW/cm$^2$ (see Appendix 1 for specific parameters). For example, when $n_4$=-4 × 10$^{-4}$ cm$^4$/TW$^2$, the optimal self-compression was achieved at an input intensity of approximately 6.4 TW/cm$^2$ resulting in a $F_{ECF}$ value of 1.5. In the case of fused silica glass plates, it has been reported that the damage threshold for kHz repetition rates laser is around 8.64 TW/cm$^2$ [37]. Consequently, the compression ratio is only 1.35, disregarding any losses even in the presence of self-compression.

*3.2 Negative pre-chirp*

Several previous studies have demonstrated that negatively pre-chirped input pulses can lead to self-compression experimentally in normally dispersive media, whereas those without pre-chirp fail to achieve the same outcome even with identical input parameters[21, 24]. These experimental findings suggest a crucial role of negative pre-chirp in self-compression in normally dispersive media. Therefore, building upon previous experiments, this section aims to thoroughly investigate the influence of pre-chirp in detail.

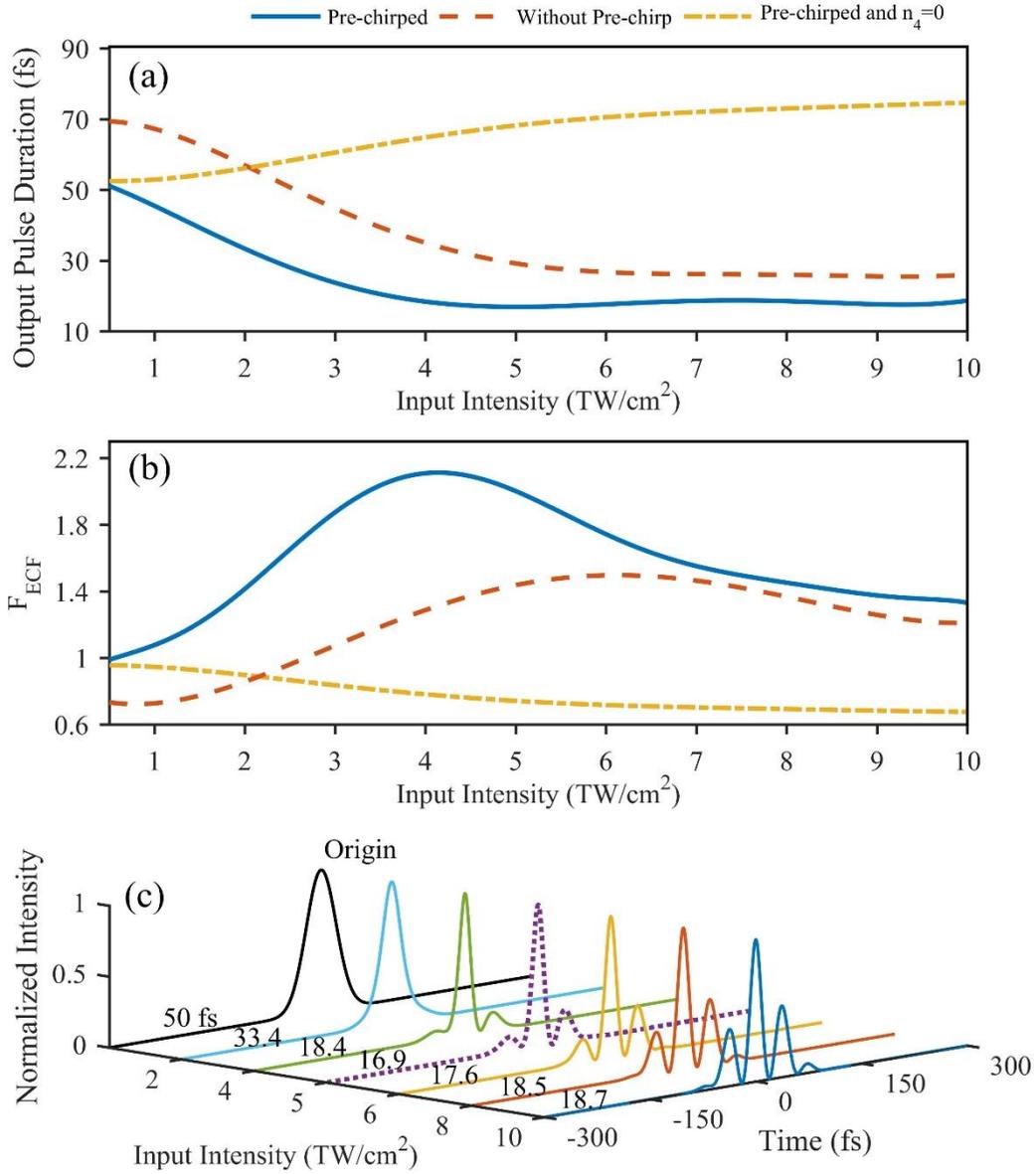

Fig.3. (a) and (b) the main output pulse durations and $F_{ECF}$ at different input intensities with and without negatively pre-chirp, respectively. Yellow dash-dotted line: $n_4$=0; red-dashed line: without pre-chirp; blue-solid line: negatively chirped. (c) Normalized temporal profile of the output pulses with 2.0 to 10.0 TW/cm$^2$ input, and 75 fs negatively pre-chirped input pulses.

By employing the identical simulated parameters as described in the preceding sections, but including an additional negatively pre-chirped 75 fs input pulse, the output durations at different input intensities are illustrated in Fig. 3(a). In the case of a negatively pre-chirped input, the output durations exhibit a continuous decrease with increasing input intensities. Furthermore, self-compression can achieve shorter pulse duration under lower input intensity compared to TL pulse input. The output duration reaches the initial TL duration of 50 fs at approximately 0.5 TW/cm$^2$. When the input intensity is raised to 5.0 TW/cm$^2$, Fig. 3(c) demonstrates that the output duration reaches its shortest duration of about 16.9 fs with small sidelobes. As the input intensity increased from 5.0 TW/cm$^2$ to 10.0 TW/cm$^2$, the output duration of the central main pulse slightly broadened, while both sidelobes exhibited a rapid increase. Additionally, Fig. 3(b) demonstrated that for negatively pre-chirped input pulses, the $F_{ECF}$ is 2.11 at 4.2 TW/cm$^2$ input intensity. In contrast, for input pulses without pre-chirp, the maximum $F_{ECF}$ is achieved at a higher input intensity of 6.1 TW/cm$^2$ with a value of 1.50. Regarding the pre-chirped input pulse with $n_4$=0, the output duration increased continuously, when $F_{ECF}$ exhibited a slow decrease simultaneously as the input intensity increasing.

The contribution of negatively pre-chirped pulses to the self-compression process can be expressed as following. For the same input peak power, negatively chirped pulses with broadened duration contain more pulse energy compared to TL pulses. When negatively pre-chirped pulses propagate in a normally dispersive medium, the positive dispersion compensates for the negative pre-chirp, resulting in pulse compressing rather than broadening during the primary process in the medium. This gradual reduction in pulse duration leads to a continuous increase in laser intensity until self-compression is triggered by the fifth-order susceptibility. During this process, the negatively pre-chirped pulse's spectra distribution will be shaped by SPM effect[24], which is directly related to the pulse TL duration. In the absence of negative pre-chirp, at the same relatively low input intensity, the input pulses fail to attain a sufficiently high intensity for utilization of the fifth-order susceptibility. Ultimately the third-order susceptibility based SPM becomes the predominant nonlinear effect responsible for pulse broadening.

In conclusion, based on the aforementioned two sections, it can be inferred that a sufficiently large negative fifth-order susceptibility is a vital prerequisite for self-compression in normally dispersive region of materials. Considering finding appropriate materials with high fifth-order susceptibility is not easy, the incorporation of negative pre-chirp becomes crucial as it facilitates self-compression with higher $F_{ECF}$ at a relatively low input intensity. This approach not only mitigates potential damage to nonlinear media but also enables self-compression in normally dispersive region of materials.

*3.3 Optimizing input parameters*

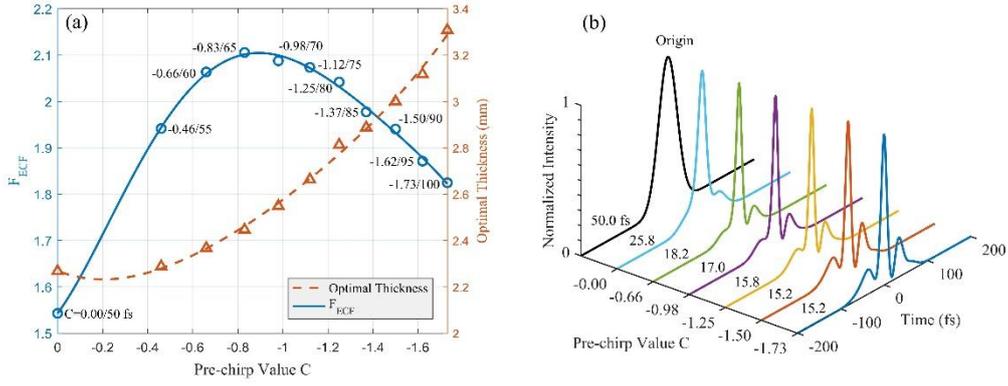

Fig. 4. (a) The maximal $F_{ECF}$ and corresponding optimal medium length changing with different pre-chirp value at 5.0 TW/cm$^2$ input intensity. (b) The corresponding normalized temporal profile of the output pulses.

To determine the optimal pre-chirp in a self-compression process, simulations were conducted to measure the maximum $F_{ECF}$ at a constant input intensity but varying levels of pre-chirp. For 5.0 TW/cm$^2$ input intensity, the maximal $F_{ECF}$ increased from 1.54 to 2.11 as the value of pre-chirp changed from -0.00 to -0.83, as illustrated in Fig. 4(a). Additionally, the optimal thickness of the glass plate required to obtain the maximal $F_{ECF}$ increased with higher absolute values of negative pre-chirp due to longer propagation length needed in the medium for attaining minimum output duration. The corresponding output temporal profiles at the maximal $F_{ECF}$ are shown in Fig. 4(b), where the output duration decreases from 25.8 fs to 15.2 fs as the absolute value of pre-chirp is increased to 1.50. However, when the absolute value of pre-chirp is further increased from 1.50 to 1.73, the decrease in output pulse duration stops due to the emergence of satellite pulses with a decreased $F_{ECF}$ from 1.94 to 1.82. Furthermore, Fig. 4(a) and Fig. 3(c) demonstrate a wide range of stable self-compression achieved with little change in $F_{ECF}$, spanning from -0.70 to -1.10 for pre-chirp values and an input intensity range of approximately 3.5 to 5.0 TW/cm$^2$. This indicates that stable self-compression can be obtained at these ranges, which are not sensitive to both the negatively pe-chirp value and the input intensity.

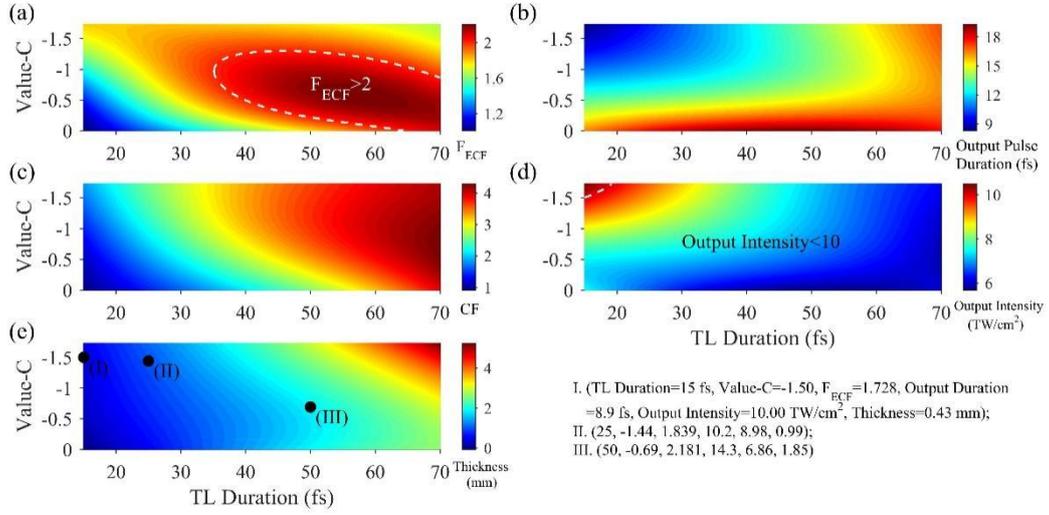

Fig. 5. (a) is the maximal $F_{ECF}$ changed with different input TL pulse durations from 15 fs to 70 fs, and different pre-chirp values from 0.00 to -1.73, at the same 7.0 TW/cm² input intensity. (b)-(e) are the corresponding output duration, CF, output intensity and medium thickness.

Besides the value of $n_4$, negative pre-chirp, and input intensity, the input TL pulse duration also plays an important role in determining the self-compression ratio. Fig. 5(a)-(e) illustrates the variations in maximal $F_{ECF}$, output duration, CF, output intensity, and thickness of fused silica plate for different input TL pulse durations and pre-chirp values under a fixed input intensity of 7.0 TW/cm², which is considered appropriate for each specific TL pulse duration. To achieve maximal $F_{ECF}$, shorter TL pulse duration requires a relatively larger pre-chirp value and a corresponding thinner media, as depicted in Fig 5(a) and (e). This is attributed to the broader spectra of shorter TL pulse, which makes them more sensitive to material-induced second-order dispersion and thus benefits from a thinner glass plate when combined with a larger absolute pre-chirp value for enhancing nonlinear effect. Conversely, longer TL pulse duration necessitate a relatively smaller pre-chirp value but thicker media to attain the maximum $F_{ECF}$. Here, the maximal $F_{ECF}$ is limited by the damage threshold of the material. Extensive research has been conducted on the damage threshold of fused silica, as detailed in Appendix 2. For 800 nm, 50 fs repetition pulses, the damage threshold intensity is about 10.0 TW/cm² in a fused silica glass[37]. However, for shorter TL pulse durations, the output intensity may exceed the damage threshold, as depicted in Fig. 5(d). As TL duration increases, self-defocusing caused by $n_4$ and multi-photon absorption in thicker media lead to a decrease in output intensity. To obtain applicable self-compression results, several certain limits are set to restrict the input parameters. The region encompassed by the dashed line in Fig. 5(a) satisfies $F_{ECF} > 2$, while in Fig. 5(d), it meets the condition of output intensity being below the damage threshold. For a typical 50, 25, 15 fs TL pulse and 7.0 TW/cm² input intensity, the optimal parameters are labeled in Fig. 5(e).

In an actual experiment, the spatial parameters of input beam also have significant influences on pulse self-compression. A thicker medium is more susceptible to the induction of SSSF and damage the medium, as a result of both spatial intensity modulation and wavefront aberration at high spatial frequencies of the input beam. Beam smoothing by inducing spatial dispersion after the grating-based compressor helps to address this problem[25, 26]. Similar to pulse compression by employing multiple thin glass plates[19, 38], it may be feasible to achieve self-compression with multiple thin plates under appropriate input parameters in the future.

## 4. Proof-of-principle experiment

A proof-of-principle experiment was conducted to validate the previous numerical conclusion, and the detailed procedure is described in the methodology section. The 30 fs TL pulses with repetition rates of 1-kHz were negatively pre-chirped to 40 fs, 46 fs and 54 fs with variational input intensities before being directed on a 2-mm-thick fused silica plate. The output spectral widths of the 46 fs pre-chirped pulse, calculated based on the $1/e^2$ of the maximum spectral intensity, are presented in Fig. 6(a). The corresponding output spectrum are clearly depicted in Fig. 6(b). The output spectral width increased from 55 nm to 68 nm as the input intensity was raised from 0 to 2.50 TW/cm$^2$. However, at an input intensity of 4.16 TW/cm$^2$, the spectral width decreased to 47 nm while pulse maintaining a shortened duration to 17.3 fs. It is inexplainable for this peculiar phenomenon that decreases in frequency and time domain widths happened synchronously, disregarding the exist of fifth-order susceptibility. Because nothing but negative fifth-order susceptibility could transform the positive frequency chirp caused by SPM into negative, generating the narrowing in spectra and compression in time domain. The temporal profiles of the input and output pulses were measured using a home-made Transient-Grating based Self-reference Spectral Interferometry (TG-SRSI)[39, 40]. Fig. 6(c) displays the retrieved input and output temporal profiles, as well as the TL temporal profile and phase of one representative output pulse. The self-compressed output pulse exhibits a FWHM duration of 17.3 fs, which closely matches its TL duration of 17.2 fs. Additionally, Fig. 6(d) illustrates the spectrum of the input pulse, output test pulse, obtained TG reference pulse, and the interference spectra between the output test pulse and TG reference pulse.

  The variation of spectrum and pulse duration in experiment clearly demonstrates that INF induced by $n_4$ and SPM is the essence resulting in spectral shaping and self-compression, which is consistent with the previous numerical analysis. Specifically, when the input pulse was pre-chirped to 40 fs and 54 fs, the output pulses were self-compressed to 23.3 and 19.7 fs, respectively. Further detailed properties and analyses of spectra and pulse profiles varied with the increasing of input intensity are given in supplement 3. Importantly, experimental results validate the conclusion drawn from our simulation analysis.

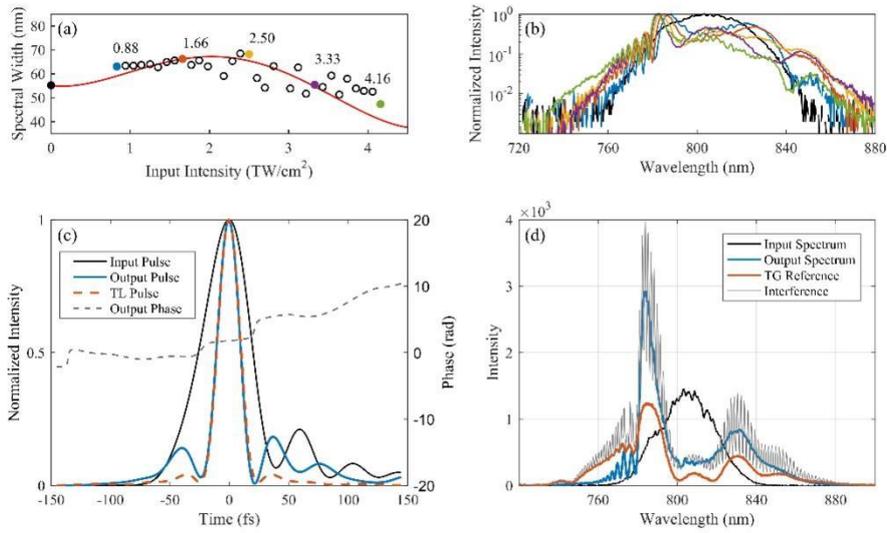

Fig. 6. (a)The output spectral widths at different input intensities with 46 fs chirped input pulses. The red-solid-line is corresponding simulation curves. (b)The corresponding normalized spectral profiles in logarithmic coordinate. (c)The retrieved input and output temporal profiles, TL temporal profile, and phase of one typical output pulse. (d)The spectra of the input pulse, the output test pulse, the obtained TG reference, and the interference spectrum by TG-SRSI.

## 5. Self-compression for PW laser

Due to the current difficulty in obtaining a thin fused silica glass plate with large size and nice optical quality currently, here we only simulate the expected self-compressed outputs using the corresponding parameters in PW systems. As for PW lasers, which typically operate at low repetition rate or single-shot mode, the damage threshold of glass plate is relatively higher based on available references and Appendix 2. For a PW laser emitting pulses at 800 nm and 25 fs TL duration, it is possible to achieve self-compression down to 9.9 fs with an input intensity of 6.9 TW/cm$^2$, as depicted in Fig. 7(a). Figure 7(b) also illustrates that the corresponding spectral bandwidth is gradually broadened as the increasing of the input intensity. Broader spectrum appears in the long wavelength region because of the self-steepening at the leading edge of the pulse. TABLE 1 shows the detail and other optimum parameters.

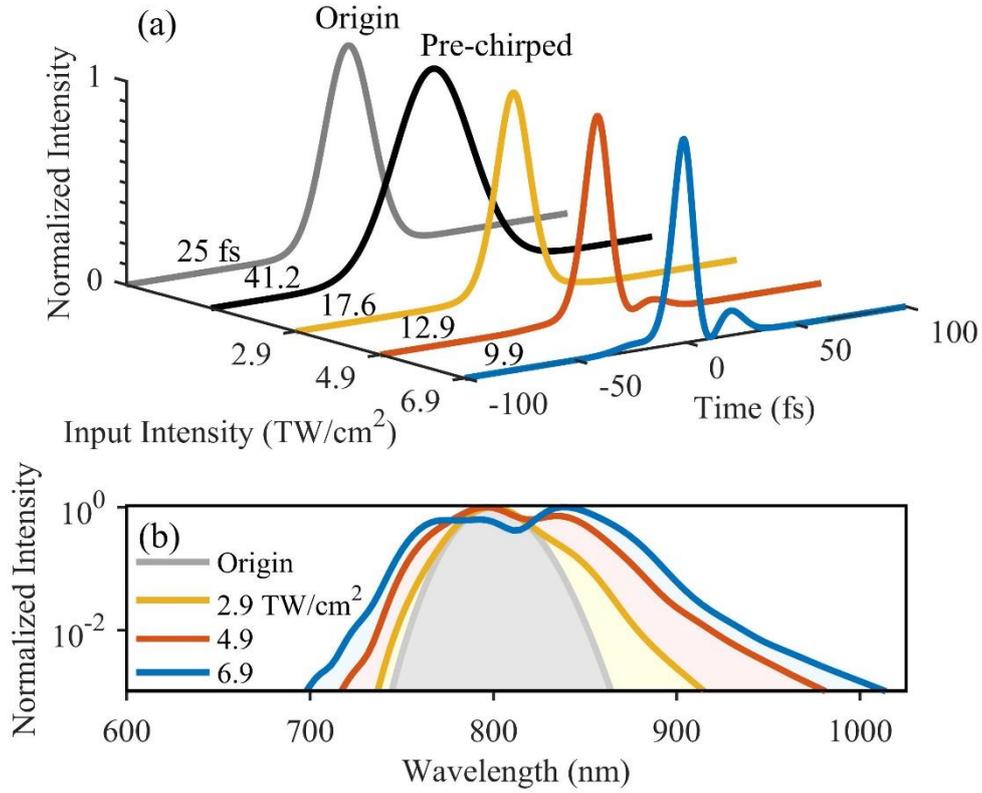

Fig. 7. (a) The normalized temporal profiles and (b) the corresponding normalized spectral profiles of the input pulse and the output pulses at different input intensities.

The simulated results and optimum input parameters for several typical ultrahigh-peak-power lasers are presented in TABLE 1, demonstrating detailed information. In order to ensure practicality, all output intensities are constrained below the damage threshold of fused silica glass at 20.00 TW/cm$^2$, while maintaining a $Q_{out}/Q_{TL\_in}$ ratio larger than 0.67. Specifically, for laser pulses at 800 nm with an input TL duration of 25 fs, even the shortest pulse duration achievable is found to be 9.88 fs, the maximum $F_{ECF}$ to be 1.96 due to limitation imposed by the output damage threshold. For a 15 fs pulse, a high $F_{ECF}$ laser pulse can be obtained with larger pre-chirp value and lower input intensity to maintain output intensity within the damage threshold. In the case of 15 fs and 25 fs laser pulses at 900 nm, thicker media are required to achieve high $F_{ECF}$ due to the smaller group velocity dispersion (GVD), compared to the 800 nm pulse. For 50 fs TL pulses at both 800 and 1030 nm, the self-compressed pulse durations close to 16 fs can be achieved with relatively thicker plates, which are restricted by $Q_{out}/Q_{TL\_in}$. It is worthing noting that we have simulated the propagation process of 1030 nm/100 fs pulses and 1064 nm/300 fs pulses under a wide range of parameters. However, due to their relatively broad input TL pulse duration, their temporal profile would ultimately broaden instead of self-compress. This is because the INF induced by $n_4$ is positively correlated with the derivative of the I$^2$ with respect to time, as demonstrated in equation 2, which means long pulse with small temporal slope cannot motivate the effect of $n_4$.

<div align="center">TABLE I. Simulation results for PW laser.</div>

| Wavelength (nm), TL duration (fs) | $n_2$ ($10^{-20}$ m²/W) | GVD ($10^{-26}$ s²/m) | Input/ Output Intensity (TW/cm²) | Value-C | Thickness (mm) | $F_{ECF}$ | Duration (fs)/CF | $Q_{out}/Q_{TL\_in}$ |
|---|---|---|---|---|---|---|---|---|
| 800,50 | 3.20[a] | 3.62 | 4.82/16.22 | -1.12 | 2.72 | 2.06 | 16.17/3.09 | 0.67 |
| 800,25 | 3.20 | 3.62 | 6.91/20.00 | -1.31 | 0.96 | 1.96 | 9.88/2.53 | 0.77 |
| 800,15 | 3.20 | 3.62 | 6.72/20.00 | -2.30 | 0.62 | 1.72 | 7.64/1.96 | 0.87 |
| 910,25 | 3.20 | 2.77 | 7.63/20.00 | -1.38 | 1.26 | 1.96 | 9.44/2.65 | 0.74 |
| 910,15 | 3.20 | 2.77 | 7.94/20.00 | -2.18 | 0.77 | 1.71 | 7.31/2.05 | 0.83 |
| 1030,50 | 2.74[b] | 1.90 | 5.13/11.89 | -1.01 | 4.92 | 1.96 | 17.01/2.94 | 0.67 |

[a]From Phys. Rev. Lett. **87**, 4 (2001).

[b]From Appl. Optics **37**, 546-550 (1998).

According to above simulations, it is feasible to effectively self-compress PW laser pulses by approximately 2 times in a thin glass plate. In combination of beam smoothing method based on asymmetric four-grating compressor[25, 26], an enhanced multistep pulse compressor for PW laser together with self-compression is shown in. The setup is simple and does not affect any optics of the laser system as the plate is positioned after the parabolic mirror.

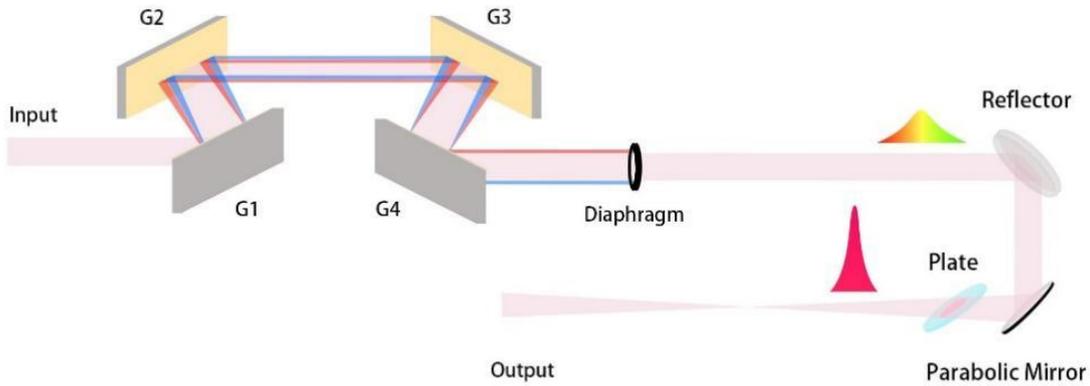

Fig. 8. Self-compression setup of PW laser together with an asymmetric four-grating compressor.

To further verify the self-compression effect using the experimental setup depicted in Fig. 8, another proof-of-principle experiment was conducted employing a small spatial size Super-Gaussian beam as Fig. 8 shown. About 7.0 mm spatial dispersion was introduced to a 30-mm-diameter input Super-Gaussian beam by using an asymmetric four-grating compressor. The input pulse, negatively pre-chirped to 46 fs from a 31 fs TL duration, contained 3.8 mJ energy. The laser pulse was self-compressed to 22.5 fs (TL 21 fs) in two 2-mm-thick fused silica plates at 5.4 TW/cm² input intensity. The input and output pulse temporal profile and spectrum are shown in Fig. 9(a) and (b) respectively. The detailed description of this experiment is shown in the method section. Due to SSSF, the hot spots or strong spatial intensity modulations appear in the input and output laser beam, which may damage the optical elements and prevent the application of self-compression. By implementing an asymmetric four-grating compressor, the SSSF is suppressed with relatively smoothing output beam. As for PW laser, much larger spatial dispersion can be induced to further enhance the smoothness of the input beam. Consequently, this approach offers a novel and practical direction to further improve the peak-power of PW or even EW laser pulse.

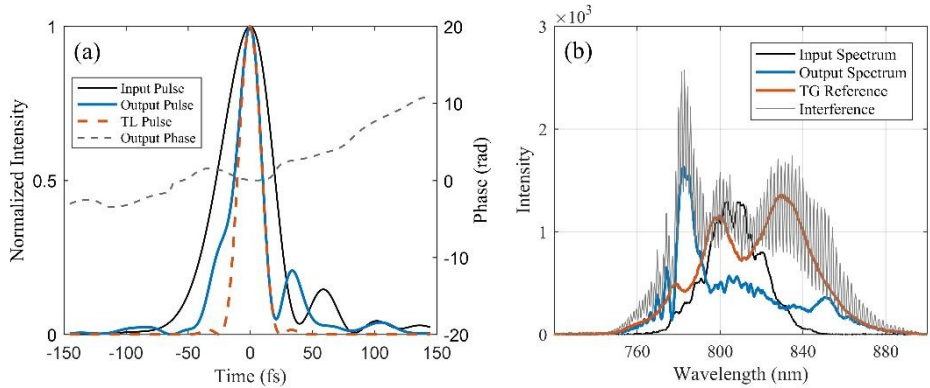

Fig. 9. (a)The retrieved input and output temporal profiles, TL temporal profile, and phase of one typical output pulse. (b)The spectra of the input pulse, the output test pulse, the obtained TG reference, and the interference spectrum by TG-SRSI.

## 6. Conclusion

The possibility of self-compression in normally dispersive media has been theoretically demonstrated, presenting a straightforward, cost-effective, and competitive approach to enhance laser peak power. To understand the self-compression process comprehensively, several variable parameters such as $n_4$, negative pre-chirp, the TL pulse duration, and thickness of medium, were considered and simulated synthetically. Two significant conclusions were derived:1) the existence of the fifth-order susceptibility is imperative for achieving self-compression in media with normal dispersion; 2) the incorporation of negative pre-chirp facilitates pulse self-compression, resulting in higher $F_{ECF}$, even at a relatively low input intensity that mitigate potential damage to the medium. Few-cycle laser pulses with compressed TL duration can be obtained at optimal parameters. For instance, PW laser pulses with input TL 25 fs at 800 nm can be self-compressed to 9.9 fs in an around 1.0-mm-thick glass plate. In combination of beam smoothing method based on asymmetric four-grating compressor[25, 26], even 10s-100s PW or EW laser may further improve its peak-power by using this simple self-compression method in the future with large and high performance thin glass plates or dielectric coated reflective mirrors.

## 7. Methods

### 7.1 Proof-of-principle experiment

The input pulse was generated by a Ti: sapphire CPA system with 1 kHz repetition rates (Coherent Legend Elite-Ultrafast Amplifier Laser Systems). The negatively pre-chirp of the input pulse was induced by tuning the distance of the grating pair in CPA compressor. Subsequently, the pre-chirped pulse was focused by a plano-convex lens (f=1000 mm) onto a 2-mm-thick fused silica plate which placed about 60 mm behind the focal point. The central part was selected out through a diaphragm, and collimated by the same plano-convex lens. The output spectra were recorded by using a spectrometer (USB2000, Ocean Optics). The output laser beam was expanded by a 6× beam expander to monitor the temporal profiles by using a home-made Transient Grating based Self-reference Spectral Interferometry (TG-SRSI)[39, 40].

### 7.2 Self-compression experiment with smoothed Super-Gaussian beam

The Super-Gaussian beam was simply induced by filtering the margin of the Gaussian beam generated by the same laser system of previous experiment without compression. An asymmetric four-grating compressor with four home-made gold-coated 1480 lines/mm gratings was used to induce suitable negative chirp and spatial dispersion. The negatively pre-chirped pulse was focused by a plano-convex lens (f=1500 mm) on a 2-mm-thick glass window, located about 80 mm after the focal point, of a 500-mm-long vacuum tube. Another 2-mm-thick glass plate was placed 10 mm after the window plate for further self-compression. The output pulse was measured with the same instruments of previous experiment.

**Reference**

1. A. Shaykin, V. Ginzburg, I. Yakovlev, A. Kochetkov, A. Kuzmin, S. Mironov, I. Shaikin, S. Stukachev, V. Lozhkarev, A. Prokhorov, and E. Khazanov, "Use of KDP crystal as a Kerr nonlinear medium for compressing PW laser pulses down to 10 fs," High Power Laser Sci. Eng. **9**, 7 (2021).

2. V. Ginzburg, I. Yakovlev, A. Zuev, A. Korobeynikova, A. Kochetkov, A. Kuzmin, S. Mironov, A. Shaykin, I. Shaikin, E. Khazanov, and G. Mourou, "Fivefold compression of 250-TW laser pulses," Phys. Rev. A **101**, 6 (2020).

3. S. Toth, R. S. Nagymihaly, I. Seres, L. Lehotai, J. Csontos, L. T. Toth, P. P. Geetha, B. Kajla, D. Abt, V. Pajer, A. Farkas, A. Mohacsi, A. Borzsonyi, and K. Osvay, "Single thin-plate compression of multi-TW laser pulses to 3.9 fs," Opt. Lett. **48**, 57-60 (2023).

4. A. Baltuska, Z. Y. Wei, M. S. Pshenichnikov, and D. A. Wiersma, "Optical pulse compression to 5 fs at a 1-MHz repetition rate," Opt. Lett. **22**, 102-104 (1997).

5. C. Jocher, T. Eidam, S. Hadrich, J. Limpert, and A. Tunnermann, "Sub 25 fs pulses from solid-core nonlinear compression stage at 250 W of average power," Opt. Lett. **37**, 4407-4409 (2012).

6. M. Nisoli, S. DeSilvestri, and O. Svelto, "Generation of high energy 10 fs pulses by a new pulse compression technique," Appl. Phys. Lett. **68**, 2793-2795 (1996).

7. T. Nagy, V. Pervak, and P. Simon, "Optimal pulse compression in long hollow fibers," Opt. Lett. **36**, 4422-4424 (2011).

8. J. E. Beetar, S. Gholam-Mirzaei, and M. Chini, "Spectral broadening and pulse compression of a 400 mu J, 20 W Yb:KGW laser using a multi-plate medium," Appl. Phys. Lett. **112**, 5 (2018).

9. J. I. Kim, Y. G. Kim, J. M. Yang, J. W. Yoon, J. H. Sung, S. K. Lee, and C. H. Nam, "Sub-10 fs pulse generation by post-compression for peak-power enhancement of a 100-TW Ti:Sapphire laser," Opt. Express **30**, 8734-8741 (2022).

10. J. Weitenberg, A. Vernaleken, J. Schulte, A. Ozawa, T. Sartorius, V. Pervak, H. D. Hoffmann, T. Udem, P. Russbuldt, and T. W. Hansch, "Multi-pass-cell-based nonlinear pulse compression to 115 fs at 7.5 mu J pulse energy and 300 W average power," Opt. Express **25**, 20502-20510 (2017).

11. M. Kaumanns, V. Pervak, D. Kormin, V. Leshchenko, A. Kessel, M. Ueffing, Y. Chen, and T. Nubbemeyer, "Multipass spectral broadening of 18 mJ pulses compressible from 1.3 ps to 41 fs," Opt. Lett. **43**, 5877-5880 (2018).

12. V. Shumakova, P. Malevich, S. Alisauskas, A. Voronin, A. M. Zheltikov, D. Faccio, D. Kartashov, A. Baltuska, and A. Pugzlys, "Multi-millijoule few-cycle mid-infrared pulses through nonlinear self-compression in bulk," Nat. Commun. **7**, 6 (2016).

13. T. Balciunas, C. Fourcade-Dutin, G. Fan, T. Witting, A. A. Voronin, A. M. Zheltikov, F. Gerome, G. G. Paulus, A. Baltuska, and F. Benabid, "A strong-field driver in the single-cycle regime based on self-compression in a kagome fibre," Nat. Commun. **6**, 7 (2015).
14. J. Y. Qian, Y. J. Peng, Y. Y. Li, B. J. Shao, Z. Liu, W. K. Li, R. Y. Feng, L. Y. Shen, Y. X. Leng, and R. X. Li, "Few-cycle mid-infrared laser based on nonlinear self-compression in solid thin plates," Opt. Lett. **46**, 5075-5078 (2021).
15. M. Hemmer, M. Baudisch, A. Thai, A. Couairon, and J. Biegert, "Self-compression to sub-3-cycle duration of mid-infrared optical pulses in dielectrics," Opt. Express **21**, 28095-28102 (2013).
16. S. Y. Mironov, I. B. Mukhin, V. V. Lozhkarev, A. K. Potemkin, M. A. Martyanov, I. V. Kuzmin, and E. A. Khazanov, "Temporal compression of high-power IR laser pulses in a KDP crystal," Appl. Optics **61**, 6033-6037 (2022).
17. A. V. Mitrofanov, A. A. Voronin, M. V. Rozhko, D. A. Sidorov-Biryukov, A. B. Fedotov, A. Pugzlys, V. Shumakova, S. Alisauskas, A. Baltuska, and A. M. Zheltikov, "Self-compression of high-peak-power mid-infrared pulses in anomalously dispersive air," Optica **4**, 1405-1408 (2017).
18. U. Elu, M. Baudisch, H. Pires, F. Tani, M. H. Frosz, F. Kottig, A. Ermolov, P. S. Russell, and J. Biegert, "High average power and single-cycle pulses from a mid-IR optical parametric chirped pulse amplifier," Optica **4**, 1024-1029 (2017).
19. V. N. Ginsburg, I. V. Yakovlev, A. S. Zuev, A. P. Korobeynikova, A. A. Kochetkov, A. A. Kuz'min, S. Y. Mironov, A. A. Shaykin, I. A. Shaikin, and E. A. Khazanov, "Compression after compressor: threefold shortening of 200-TW laser pulses," Quantum Electron. **49**, 299-301 (2019).
20. Z. Y. Li, Y. Kato, and J. Kawanaka, "Simulating an ultra-broadband concept for Exawatt-class lasers," Sci Rep **11**, 16 (2021).
21. A. I. Aristov, Y. V. Grudtsyn, L. D. Mikheev, A. V. Polivin, S. G. Stepanov, V. A. Trofimov, and V. I. Yalovoi, "Spectral broadening and self-compression of negatively chirped visible femtosecond pulses in fused silica," Quantum Electron. **42**, 1097-1099 (2012).
22. J. C. Travers, T. F. Grigorova, C. Brahms, and F. Belli, "High-energy pulse self-compression and ultraviolet generation through soliton dynamics in hollow capillary fibres," Nat. Photonics **13**, 547-+ (2019).
23. O. Shorokhov, A. Pukhov, and I. Kostyukov, "Self-compression of laser pulses in plasma," Phys. Rev. Lett. **91**, 4 (2003).
24. J. Liu, X. W. Chen, J. S. Liu, Y. Zhu, Y. X. Leng, J. Dai, R. X. Li, and Z. Z. Xu, "Spectrum reshaping and pulse self-compression in normally dispersive media with negatively chirped femtosecond pulses," Opt. Express **14**, 979-987 (2006).
25. J. Liu, X. Shen, S. M. Du, and R. X. Li, "Multistep pulse compressor for 10s to 100s PW lasers," Opt. Express **29**, 17140-17158 (2021).
26. X. Shen, S. M. Du, W. H. Liang, P. Wang, J. Liu, and R. X. Li, "Two-step pulse compressor based on asymmetric four-grating compressor for femtosecond petawatt lasers," Appl. Phys. B-Lasers Opt. **128**, 14 (2022).
27. K. Ekvall, C. Lundevall, and P. van der Meulen, "Studies of the fifth-order nonlinear susceptibility of ultraviolet-grade fused silica," Opt. Lett. **26**, 896-898 (2001).
28. V. Besse, H. Leblond, and G. Boudebs, "Fifth-order nonlinear susceptibility: Effect of third-order resonances in a classical theory," Phys. Rev. A **92**, 10 (2015).

29. V. Jarutis and V. Vaicaitis, "Fifth-order nonlinear susceptibilities of sodium atomic vapor," European Physical Journal Plus **129**(2014).
30. E. L. Falcao-Filho, R. Barbosa-Silva, R. G. Sobral-Filho, A. M. Brito-Silva, A. Galembeck, and C. B. de Araujo, "High-order nonlinearity of silica-gold nanoshells in chloroform at 1560 nm," Opt. Express **18**, 21636-21644 (2010).
31. Z. B. Liu, W. P. Zang, J. G. Tian, W. Y. Zhou, C. P. Zhang, and G. Y. Zhang, "Analysis of Z-scan of thick media with high-order nonlinearity by variational approach," Opt. Commun. **219**, 411-419 (2003).
32. S. Akturk, X. Gu, P. Bowlan, and R. Trebino, "Spatio-temporal couplings in ultrashort laser pulses," J. Opt. **12**, 20 (2010).
33. S. Tzortzakis, L. Sudrie, M. Franco, B. Prade, A. Mysyrowicz, A. Couairon, and L. Berge, "Self-guided propagation of ultrashort IR laser pulses in fused silica," Phys. Rev. Lett. **87**, 4 (2001).
34. G. Shi, Y. Wang, X. Zhang, K. Yang, and Y. Song, *Investigation of high-order optical nonlinearities by the Z-scan technique*, International Symposium on Photoelectronic Detection and Imaging 2009 (SPIE, 2009), Vol. 7382.
35. P. Russbueldt, J. Weitenberg, J. Schulte, R. Meyer, C. Meinhardt, H. D. Hoffmann, and R. Poprawe, "Scalable 30 fs laser source with 530 W average power," Opt. Lett. **44**, 5222-5225 (2019).
36. A. L. Viotti, S. Alisauskas, H. Tunnermann, E. Escoto, M. Seidel, K. Dudde, B. Manschwetus, I. Hartl, and C. M. Heyl, "Temporal pulse quality of a Yb:YAG burst-mode laser post-compressed in a multi-pass cell," Opt. Lett. **46**, 4686-4689 (2021).
37. S. Z. Xu, C. Z. Yao, W. Liao, X. D. Yuan, T. Wang, and X. T. Zu, "Experimental study on 800 nm femtosecond laser ablation of fused silica in air and vacuum," Nucl. Instrum. Methods Phys. Res. Sect. B-Beam Interact. Mater. Atoms **385**, 46-50 (2016).
38. S. Y. Mironov, J. A. Wheeler, E. A. Khazanov, and G. A. Mourou, "Compression of high-power laser pulses using only multiple ultrathin plane plates," Opt. Lett. **46**, 4570-4573 (2021).
39. J. Liu, F. J. Li, Y. L. Jiang, C. Li, Y. X. Leng, T. Kobayashi, R. X. Li, and Z. Z. Xu, "Transient-grating self-referenced spectral interferometry for infrared femtosecond pulse characterization," Opt. Lett. **37**, 4829-4831 (2012).
40. Z. Si, X. Shen, J. X. Zhu, L. Lin, L. H. Bai, and J. Liu, "All-reflective self-referenced spectral interferometry for single-shot measurement of few-cycle femtosecond pulses in a broadband spectral range," Chin. Opt. Lett. **18**, 5 (2020).

**Data availability**

The data that supports the plots within this paper and other findings of this study are available from the corresponding author upon reasonable request.

**Code availability**

The simulation codes that support the findings of the study are available from the corresponding author upon reasonable request.

**Acknowledgements**

The authors wish to acknowledge Prof. E. A. Khazanov for the discussion. This work was supported by the Shanghai Municipal Natural Science Foundation (No. 20ZR1464500), the National Natural Science Foundation of China (NSFC) (Nos. 61905257 and U1930115) and the Shanghai Municipal Science and Technology Major Project (No. 2017SHZDZX02).

**Author contributions**

J. L. and R.Ch. conceived the idea. R.Ch. performed the simulations. R. Ch., W. L. and Y. X. performed the experiments. J. L., R. Ch., X. Sh., and P. W. analyzed the data. R.Ch. and J. L. prepared the manuscript and discussed with all authors. J. L. and R. L supervised the project.

**Competing interests**

The authors declare no competing financial interests.

**Figure Legends**

Fig. 1. The evolution of INF depending on time at different input intensities from 0.8 to 4.8 TW/cm$^2$ (a) without and (b) with the fifth-order susceptibility.

Fig. 2. (a) and (b) The output duration and $F_{ECF}$ curves changing with the input intensities. (c) The output normalized temporal profiles in linear coordinate with $n_4$=-4×10$^{-4}$ cm$^4$/TW$^2$. (d) The fitted curves of critical input intensities achieving self-compress or the minimum duration at different $n_4/n_2$.

Fig.3. (a) and (b) the main output pulse durations and $F_{ECF}$ at different input intensities with and without negatively pre-chirp, respectively. Yellow dash-dotted line: $n_4$=0; red dashed line: without pre-chirp; blue solid line: negatively chirped. (c) Normalized temporal profile of the output pulses with 2.0 to 10.0 TW/cm$^2$ input, and 75 fs negatively pre-chirped input pulses.

Fig. 4. (a) The maximal $F_{ECF}$ and corresponding optimal medium length changing with different pre-chirp value at 5.0 TW/cm$^2$ input intensity. (b) The corresponding normalized temporal profile of the output pulses.

Fig. 5. (a) is the maximal $F_{ECF}$ changed with different input TL pulse durations from 15 fs to 70 fs, and different pre-chirp values from 0.00 to -1.73, at the same 7.0 TW/cm$^2$ input intensity. (b)-(e) are the corresponding output duration, CF, output intensity and medium thickness.

Fig. 6. (a)The output spectral widths with different input intensities of 46 fs input pulses. The black solid lines are corresponding simulation curves. (b)The corresponding normalized spectral profiles in logarithmic coordinate. (c)The retrieved input and output temporal profiles, TL temporal profile, and phase of one typical output pulse. (d)The spectra of the input pulse, the output test pulse, the obtained TG reference, and the interference spectrum by TG-SRSI.

Fig. 7. (a) The normalized temporal profiles and (b) the corresponding normalized spectral profiles of the input pulse and the output pulses at different input intensities.

Fig. 8. Self-compression setup of PW laser together with an asymmetric four-grating compressor.

Fig. 9. (a)The retrieved input and output temporal profiles, TL temporal profile, and phase of one typical output pulse. (b)The spectra of the input pulse, the output test pulse, the obtained TG reference, and the interference spectrum by TG-SRSI.

**Additional information**

**Supplementary information** The online version contains supplementary material available at https://doi.org/...

**Correspondence and requests for materials** should be addressed to J. L.